\documentclass[preprint,showpacs]{revtex4}

\begin{document}
\title{Topological spin current induced by non-commuting coordinates-An application to the Spin-Hall effect }
\author{D.Schmeltzer}
\affiliation{Department of Physics,\\City College of the City University Of New York\\
New York,N,Y, 10031}
\date {\today}

\begin{abstract}

We show that the $SU(2)$ transformation which diagonalizes the two dimensional spin-orbit hamiltonian  has a singularity in the momentum space at $\vec{K} = 0$.This  gives rise to  non-commuting cartesian coordinates.  When an external electric field is applied, the non-commuting cartesian coordinates induce a Hall current.  
  The spin-Hall conductance is quantized in units of  $\frac{e g \mu_{B}}{2 h}$, and the charge-Hall conductivity controlled  by the strength of the  Zeeman magnetic field.
 We propose an experiment to measure the spin-Hall conductance.This proposal is based on a magnetic field gradient which induces an electric current proportional to  the spin-Hall conductance.  
\end{abstract}


\maketitle

\textit{Introduction}.  One possibility for generating a spin current in a two dimensional electron gas (2DEG) is based on the Aharon Casher (A.C.) effect \cite{ac,bal}, known in Condensed Matter as the spin-orbit (SO) Rashba hamiltonian \cite{rash,mol,los}.  Recently, it has been proposed that a dissipationless spin current might be possible.Such a proposal\cite{mura} has been realized  for the Luttinger hamiltonian \cite{lutt} where a Berry phase has been derived in the momentum space.  No such result with the connection to the Berry phase has been established for the 2DEG Rashba hamiltonian.  This fact has not prevented researchers to propose a dissipationless current for the Rashba hamiltonian, \cite{simova,bau,raimond,olga,halperin}.  In particular, we mention the result obtained in ref.\cite{simova} where the authors have used a mean field time dependent Bloch equation for a spin in a time dependent magnetic field to compute the spin-Hall current.  Unlike in the Quantum Hall case, this current was not based on the existence of a Berry phase.

The effect of a short range elastic scattering due to the disorder potential causes the spin-Hall current to vanish \cite{bau,halperin}. In ref.\cite{bau} the authors have shown that the elastic scattering gives rise to strong vertex corrections causing the Hall current to vanish.  In ref.\cite{halperin}, a detailed calculation, which takes into account the spin orbit splitting $\Delta$, the elastic scattering rate $\tau^{-1}$, the frequency $\omega$ and the size of the system, has been performed.  The conclusion from this calculation is that the spin-Hall conductivity vanishes in an infinite system independent of the spin-orbit splitting $\Delta$ and the elastic scattering time $\tau^{-1}$.

We will show that the Spin-Hall current for the Rashba hamiltonian is derived from a Berry phase like in the Quantum  Hall case.In particular, we show that the result obtained by Simova et al \cite{simova}  is caused by the existence of a monopole at $\vec{K}=0$ and does not depend on the two dimensional density (see eqs 5-7 in ref.\cite{simova}).  The existence of the Hall current depends only on the occupancy of the electronic state at $\vec{K} =0$.  The derivation of our result is based on the Berry phase which emerges from the diagonalization of the hamiltonian in the presence of an external electric field $E^{ext}_{1}$.  The eigenfunctions which diagonalize the Rashba hamiltonian are given by a $SU(2)$ spinor in  the momentum and spin 1/2 space.  The SU(2) spinor is represented in terms of the polar angles,$U(\vec{K})\equiv U(\theta(\vec{K}), \phi(\vec{K}))$ where $\theta$ and $\phi$ are the polar and the azymutal angles on a $S_{2}$ sphere (the surface of unit sphere in 3-dimensions).  The presence of an external electric field $E^{ext}_{1}$ gives rise to a time dependent behavior of the quasimomentum $\vec{K}=(K_{1},K_{2})$ in the two dimensional Brillouine zone which is equivalent to a two dimensional Torus.  Following ref.\cite{zak,kohm,niu} we find that the Berry phase originates from  the  connection (vector potential), $\vec{A}(\vec{K})=i U^{\dagger}(\vec{K})\frac{\vec{\partial}}{\partial K}U(\vec{K})$ which has a non zero curvature caused by the azymutal multivalued function  $\phi( \vec{K})$.
  The Hall current will be computed within a new method which focuses on the the non-dissipative current (for the dissipative part we have to follow the methodology given in refs. \cite{bau,raimond,olga,halperin}).  We construct the velocity operator in the presence of an external electric field.The explicit derivation is based on the $SU(2)$ transformation which diagonalizes the Rashba hamiltonian \cite {bau,raimond,olga}.  Performing this transformation in the presence of the external electric field, we discover that the transformation has a singularity at $\vec{K}=0$ which gives rise to a monopole and therefore to a spin-Hall current.  We mention that we are not able to comment on the calculation given in ref.\cite {halperin}, since the  authors have not used the $SU(2)$ transformation to diagonalize the hamiltonian.  The Hall current will be computed from the expectation value with respect to a wave function which depends on the disorder potential, but is independent of the external electric field (the electric field has been included in the current operator).  The Hall current depends on the ground state properties of a disordered system and is sensitive to the occupation of the single particle state at $\vec{K}=0$.
         
We find that the spin-Hall conductance is quantized in units of $\frac{e g \mu_{B}}{2 h}$ (where $e$ is the electric charge, $\frac{g}{2}\mu_{B}$ is the magnetic charge and $h$ is the Planck constant).  When the magnetic Zeeman interaction is present, we have in addition an anomalous charge-Hall current.  For a two dimensional electron gas with a finite width in the $i=3$ direction, we find that the  spin Hall conductance is given by $N(\frac{e g\mu_{B}}{2 h})$, where $N$ is the number of occupied bands.
We  propose an experiment to measure the spin Hall conductance .When an external magnetic in the $i=3$ direction has a linear gradient in the $i=2$ direction  an electric current proportional to the spin Hall conductance will flow in the $i=1$ direction. 

 The plan of this paper is as following: In the first part we present the emergences of the \textit{non-commuting  cartesian coordinates} for the spin orbit problem.  In the second part, we construct a formula for the  \textit{non-dissipative Hall current}.  The last part is devoted to applications. 
  We limit ourselves to a finite system and compute the spin and charge Hall current for \textit{zero and a non-zero magnetic field} and investigate  the  effect of a finite \textit{width} $D_{z}$  in the $i=3$ direction.  
 
\textit{The model}. We consider the   one band spin-orbit hamiltonian in the Kohn-Luttinger momentum representation in the presence of a  static disorder potential and a Zeeman interaction term
\begin{equation} 
h(K)=h_{o}(K)+  V_{d}(i\frac{\vec{\partial}}{\partial K}) 
\label{hamiltonian}
\end{equation}
where $h_{o}(K)$ is the one body hamiltonian used in refs.\cite{bau,raimond}, and
\begin{equation}
h_{o}(K)=\frac{\hbar^{2}(\vec{K}-{K}_{so}(\vec{\sigma}\times\hat{e}_{3}))^{2}}{2m^{\ast}}+ \frac{1}{2}g \mu_{B}\sigma_{3}B - e \tilde{E}^{ext}_{1}(\frac{i \partial}{\partial K_{1}})
\label{homiltonian}
\end{equation}
where $K_{so}=\frac{g\mu_{B}}{2c\hbar}E_{0}$ is the spin-orbit momentum defined in terms of the magnetic charge and the effective intrinsic electric field $E_{0}$ \cite{rash} directed in the $\hat{e_{3}}$ direction (perpendicular to the two dimensional electron gas), and $\vec{\sigma}$ is the Pauli matrix.  The second term is the Zeeman term with the magnetic field B in the $i=3$ direction, $V_{d}(i\frac{\vec{\partial}}{\partial K})$ is the random potential, $\tilde{E}_1^{ext} = \vartheta(t) E_1^{ext}$ is the probing electric field in the $i=1$ direction, which is non-zero for $t>0$ ($\vartheta(t)$ is the temporal step function), and $\vec{r}=i\frac{\vec{\partial}}{\partial K}$ is the coordinate in the momentum representation.

In the second quantized form, the hamiltonian is given by, $H=\sum_{K}C^{\dagger}(K)h(K)C(K)$, where $C(K)$ and $C^{\dagger}(K)$ are the two component spinors in the laboratory frame with the Pauli matrix $\sigma_{3}$ perpendicular to the 2DEG.In this frame the matrix coordinate is given by $\vec{r}=I\frac{i\vec{\partial}}{\partial K}$ where $I$ is the unity matrix in the spin space. The solution of this model is based on a $SU(2)$ transformation which  changes the representation of the Cartesian coordinates.The effect of the $SU(2)$ transformation are best understood within the Lagrangian density formalism, $L=C^{\dagger}(K,t)i\hbar\partial_{t}C(K,t)-C^{\dagger}(K,t)h(K)C(K,t)$.

\textit{The model in the transformed (rotated) frame}. In the transformed frame the spinor fields $C(K)$ and $C^{\dagger}(K)$ are replaced by $C(K)=U(K)\hat{C}(K)$ and $C^{\dagger}(K)=\hat{C}^{\dagger}(K)U^{\dagger}(K)$, where $U(K)$ represents the mapping between the two reference systems.  In the transformed frame the lagrangian density $L$ is replaced by
\begin{eqnarray}
L&=&\hat{C}^{\dagger}(K,t)\left[i\hbar\partial_{t}+i\hbar U^{\dagger}(K)\partial_{t}U(K)\right]\hat{C}(K,t)\nonumber\\ &-& \hat{C}^{\dagger}(K,t)\left[\hat{h}_{o}(K)+V_{d}(\hat{r}(K))- e \tilde{E}^{ext}_{1}\hat{r}_{1}(K)\right]C(K,t)
\label{lagrangian}
\end{eqnarray}
where $\hat{h}_{o}(K)=U^{\dagger}(K)h_{o}(K)U(K) \equiv \hat{\hat{h}}_o(K) + e \tilde{E}_1^{ext} \hat{r}_1(K)$ is the transformed hamiltonian and 
$\hat{\vec{r}}(K)=I\frac{i\vec{\partial}}{\partial K} +U^{\dagger}(K)i\frac{\vec{\partial}}{\partial K}U(K)$ is the transformed matrix coordinates.

For a time independent transformations, we have $U^{\dagger}(K)\partial_{t}U(K)=0$ and the hamiltonian in the second quantized form is given by
\begin{equation}
H=\sum_{K}\hat{C^{\dagger}}(K)\left[\hat{\hat{h}}_{o}(K) +V_{d}(\hat{r}(K))-e \tilde{E}^{ext}_{1}\hat{r}_{1}(K)\right]\hat{C}(K)
\label{new hamiltonian}
\end {equation}
where $\hat{\hat{h}}_{o}(K) = \hat{h}_{o}(K, {E}_1^{ext}=0)$ is the hamiltonian in the absence of the  external electric field.

\textit{The velocity operator in the transformed (rotated) frame}.  Using the Heisenberg equation of motion in the first quantized form we find,
\begin {equation}
\frac{d\hat{r}_{i}}{dt}=\frac{1}{i\hbar}\left[\hat{r}_{i}(K),\hat{\hat{h}}_{o}(K)\right]+\frac{1}{i\hbar}\left[\hat{r}_{i}(K),V_{d}(\hat{r}(K))\right]-\frac{e \tilde{E}^{ext}_{1}}{i\hbar}\left[\hat{r}_{i}(K),\hat{r}_{1}(K)\right]
\label{motion}
\end{equation}
We observe that, for a singular $SU(2)$ transformation, the commutator $[\hat{r}_{1}(K),\hat{r}_{2}(K)]$ is non-zero and gives rise to a non-dissipative Hall current induced by the electric field $E^{ext}_{1}$ (the last term in equation 5).  

\textit{The diagonalization of the hamiltonian $h_o(K)$}.  Following ref.[7,8], we find that the hamiltonian $h_o(K)$  in the absence of the external electric field is diagonalized with the help of the spinor $U(K) = U(\theta(K), \phi(K)) = \left(e^{(-i/2)\phi(K)\sigma_{3}}\right)\left(e^{ (-i/2)\theta(K)\sigma_{2}}\right)$.  The angle $\phi(K)$  represents the rotation in the $x-y$ plane and is given by the multivalued function $\phi(K)=\arctan(\frac{-K_{1}}{K_{2}})$ , the closed line integral of the phase$\oint\frac{d\phi(K)}{2\pi}=1$ around $K=0$ represents the vorticity of the transformation. $\theta(K)$ is the rotation in the $x-z$ plane given by                                 $\tan\theta(K)=\frac{\lambda_{so}K}{b_{3}}$ where $\lambda_{so}=\frac{2\epsilon_{so}}{K_{so}}$ is the spin orbit coupling with the spin orbit energy $\epsilon_{so}(K)=\frac{\hbar^{2}K_{so}^{2}}{2m^{\ast}}$, $b_{3}= \frac{1}{2} g\mu_{B}B$ is the Zeeman energy, and $K=|\vec{K}|$ is the absolute value of the momentum.  As a result of the transformation, the hamiltonian $h_{o}(K)$ is replaced by the diagonal representation $\hat{h}_{o}(K) = \hat{\hat{h}}_o(K) - e \tilde{E}_1^{ext} \hat{r}_1 (K)$, where 
\begin{equation}
\hat{\hat{h}}_{o}(K)=\epsilon_{so}+\epsilon(K)+\sigma_{3}\Delta(K)
\label{free part}
\end{equation}
where $\Delta(K)=\sqrt{b^{2}_{3}+(\lambda_{so}K)^{2}}$ and $\epsilon(K)=\frac{\hbar^{2}K^{2}}{2m^{\ast}}$ are the polarization and the kinetic energy.  
We observe that  the unitary transformation  $U(K)$ has changed the coordinate representation from $\vec{r}=i\vec{\partial}/\partial K$ to $\vec{\hat{r}}=I i\vec{\partial}/\partial K +\vec{A}(K)$. $\vec{A}(K)$ is  the SU(2) gauge field  generated by the unitary transformation $iU^{\dagger}(K)\frac{\vec{\partial}}{\partial K}U(K)$ and is given by,
\begin{equation}
\vec{A}(K)=\frac{1}{2}[-\sigma_{1}sin\theta(K)\frac{\vec{\partial}\phi(K)} {\partial K} +\sigma_{2}\frac{\vec{\partial}\theta(K)}{\partial K} +\sigma_{3}\cos\theta(K)\frac{\vec{\partial}\phi(K)}{\partial K}]               \label{gauge field}
\end{equation}
Similarly, the Pauli matrices $\sigma_{i}$ are replaced by the transformed Pauli matrices $\hat{\sigma_{i}}=U^{\dagger}(K)\sigma_{i} U(K)$,$i=1,2,3$.

\textit{The charge-spin-Hall current}.  We will limit our investigation to the Hall currents.  In this case, we introduce a simple representation which takes advantage of the fact that the Hall current is determined by the $non$-$commuting$ $coordinates$ $\left[ \hat{r}_1(K), \hat{r}_2
(K) \right] \neq 0$.  Using the explicit form of the term $U^{\dagger}(K)i\frac{\vec{\partial}}{\partial K}U(K)$ given in eq .7 allows us to compute the $commutator$ for the $transformed$ $coordinates$,$\hat{r}_{1}$ and $\hat{r}_{2}$.  We obtain
\begin{equation}
\left[\hat{r}_{1}(K),\hat{r}_{2}(K)\right]=\frac{i}{2}\left[-\sigma_{1}\sin\theta(K)+\sigma_{3}\cos\theta(K)\right]2\pi\delta^{2}(K)
\label{commutator}
\end{equation}
This result follows from the equations $(\partial_{1}\partial_{2}-\partial_{2}\partial_{1})\theta(K)=0$ and $(\partial_{1}\partial_{2}-\partial_{2}\partial_{1})\phi(K)=2\pi\delta^{2}(K)$.  The origin of the delta function is the multivalued function $\phi(K)=\arctan(\frac{-K_{1}}{K_{2}})$ which determines the rotation in the $x-y$ plane.  (It is possible to avoid the use of one singular transformation and to derive eq.8 using two different regular transformations which differ by a gauge transformation for different regions in the momentum space.  This is similar to the Dirac monopole problem \cite{kohm,naka}.)  Eq. 8 describes a monopole of strength $ - \hbar /2$ sitting at the origin ($\vec{K} = 0$) which is the point of degeneracy of the hamiltonian in eq. 2.  The $singularity$ at $\vec{K} = 0$ gives rise to a $spin$ and $charge$ $Hall$ $current$.

\textit{The effect of disorder}.We will work in the interaction picture, where the only perturbation  hamiltonian is given by the disorder potential $V_d(\hat{r}(K))$.  The velocity operator in the interaction picture is controlled by the free hamiltonian (in the presence of the external field $\hat{h}_o(K) = \hat{\hat{h}}_o(K) - e \vartheta (t) {E}_1^{ext} \hat{r}_1(K)$).  In the interaction picture ($I$ stands for interaction picture), the velocity operator is given by
\begin{equation} \nonumber
\frac{d \hat{r}_{2, I}(K,t)}{d t} = \frac{1}{i \hbar} \left[ \hat{r}_{2, I}(K,t), \hat{h}_{o}(K,t)\right].
\end{equation}
The velocity operator  $\frac{d \hat{r}_{2, I}}{d t} $  can be decomposed into two parts:  a dissipative part given by $\frac{d\hat{r}^{(D)}_{2,I}}{d t}$=$\frac{1}{i\hbar}\left[\hat{r}_{2,I}(K,t),\hat{\hat{h}}_{o}(K,t)\right]$ with the vector $\vec{K}$ obeying the equation $\frac{d K_{1} }{d t}=-\frac{1}{i\hbar}\vartheta(t)e E^{ext}_{1}[K_{1}(t),\hat{r}_{1,I}(K,t)]$ and  a $non$-$dissipative$ part which for $t>0$ is  linear in the external field,$\frac{d\hat{r}^{(ND)}_{2,I}(K,t)}{d t}=-\frac{e E_{1}^{ext}}{i\hbar}\left[\hat{r}_{2,I}(K,t),\hat{r}_{1,I}(K,t)\right]$.  In the interaction picture, the expectation value for $t>0$ is determined by the eigenstate at $t=0$, $ | \Psi_I(0)\rangle$ , which in the absence of disorder is given by $ | \Phi_o \rangle$ (the eigenstate spinor of the hamiltonian  $\hat{\hat{h}}_o(K)$).  The existence of the non-commuting coordinates determines the Hall velocity matrix elements for $t>0$
\begin{equation}
\langle \Psi_I(0) | \frac{d \hat{r}^{(ND)}_{2, I} (K,t)}{d t} | \Psi_I(0) \rangle = - \frac{e E_1^{ext}}{i \hbar} \langle \Psi_I(0) | \left[ \hat{r}_{2,I}(K,t), \hat{r}_{1,I}(K,t) \right] | \Psi_I(0)\rangle.
\end{equation}
Eq. 10 is a result of the non-commuting coordinates given by eq. 8 .  ($For$ $a$ $periodic$ $potential$ the result given in eq.10 follows from the adiabatic theorem \cite{kohm}.)
  
 For a disorder potential, we have to compute the state $| \Psi_I (0) \rangle$.  This state is given by  a power series in the random potential $V_d(\hat{r}(K))$ which acts on the unperturbed state, $| \Psi_I(t= - \infty ) \rangle \equiv | \Phi_o (K) \rangle $.  As a result the scattered state spinor $| \Psi_I (K,0)\rangle$ is given in terms of   the unperturbed spinor $| \Phi_o (K)\rangle = |\Phi_{o,\uparrow}(K), \Phi_{o, \downarrow}(K) \rangle$.We have,
\begin{equation}
| \Psi_I(K,0) \rangle = T \left[ \exp \frac{-i}{\hbar}\int_{-\infty}^0 d t' V_{d} [ \hat{r}_{I}(K,t')] \right] | \Phi_o(K) \rangle
\end{equation}
  Using the eigenstate $| \Psi_I(K,0)\rangle$ and the velocity operator $\frac{d \hat{r}_{2, I}(K,t)}{dt}$, one can compute the ground state Hall current.We use the second-quantized form and introduce the time ordered Green's function  for the fermion operators $\hat{C}^{\dagger}(K)$ and $\hat{C}(K)$ .The ground-state expectation value of the current operators is obtained with the help of equation 7.7 page 66 in Fetter and Walecka  \cite{Fetter}.Using this formulation we compute the charge-Hall and spin-Hall current.
  
     The charge-Hall  current for $t>0$ is
\begin{eqnarray} \nonumber
J_{2, Hall}^{(c)}(t) &=& e \int \frac{d^2 K}{(2 \pi)^2} \lim_{\epsilon \rightarrow 0} \lim_{\vec{K}' \rightarrow \vec{K}} (-i) Tr \left[ I \frac{d \hat{r}_{2,I}(K,t)}{d t} \hat{G}(\vec{K},0; \vec{K}', \epsilon) \right] \\
 &=& e \int \frac{d^2 K}{(2 \pi)^2} \int \frac{d \omega}{2 \pi} (-i) Tr \left[ I \frac{d \hat{r}_{2,I}(K,t)}{d t} \hat{G}(\vec{K},\vec{K}; \omega) \right]
\end{eqnarray}
In eq.12   $e$ is the electric charge ,$I$ is the identity matrix in the spin space  and $\hat{G}$is the  time ordered Green's function in the matrix form.
 
The spin Hall current for $t>0$ with the magnetic charge $\frac{g \mu_B}{2}$ is
\begin{eqnarray} \nonumber
J_2^{(M)}(t) &=& (\frac{g}{2}\mu_B) \int \frac{d^2 K}{(2 \pi)^2} \lim_{\epsilon \rightarrow 0} \lim_{\vec{K}' \rightarrow \vec{K}} (-i) Tr \left[ \hat{\sigma}_3 \frac{d \hat{r}_{2,I}(K,t)}{d t} \hat{G}(\vec{K},0; \vec{K}', \epsilon) \right] \\
 &=& (\frac{g}{2}\mu_B) \int \frac{d^2 K}{(2 \pi)^2} \int \frac{d \omega}{2 \pi} (-i) Tr \left[\hat{\sigma_{3}} \frac{d \hat{r}_{2,I}(K,t)}{d t} \hat{G}(\vec{K},\vec{K}; \omega) \right]
 \end{eqnarray}
``Tr'' represents the trace over the spin space and $\hat{\sigma_{3}}$represents the transformed Pauli matrix in the spin space.  In eqs. 12 and 13, we have introduced the time ordered single particle Green's function $\hat{G}_{\alpha, \beta} (\vec{K}, 0; \vec{K}', \epsilon)$ with $\alpha, \beta = \uparrow, \downarrow$ being the spin indices and the transformed Pauli matrix $\hat{\sigma}_3 = U^\dag(K) \sigma_3 U(K) = \sigma_3 \cos \theta(K) - \sigma_1 \sin \theta(K)$.  The limit $\vec{K}' \rightarrow \vec{K}$ can be safely taken since the current operator depends on the momentum $\vec{K}$ but not on the momentum derivative.  The  non-dissipative Hall current is obtained when we replace the velocity operator in eqs.12,13 by the $non$-$dissipative$ velocity $\frac{d\hat{r}^{(ND)}_{2,I}(K,t)}{d t}$  (see eq.10),this formulas are similar to the   non-classical contribution obtained for the Hall effect \cite{str} which depends only on the number of occupied states below the Fermi energy. 

In order to compute the currents in equations 12,13 we need to have  the single particle Green's functions.Since the non-dissipative velocity is linear in the external electric field we can use the time ordered Green's function  for  a zero external  electric field. The time ordered green's function at $T=0$   is  a function  only of the disorder potential $V_d(\hat{r}(K))$  and obeys the integral equation
\begin{eqnarray} \nonumber
& &\hat{G}_{\alpha, \beta}(\vec{K},\vec{K}';\omega) = \hbar \delta_{\vec{K}, \vec{K}'} \delta_{\alpha, \beta} \hat{G}^{(o)}_{\alpha, \alpha}(\vec{K}, \vec{K}'; \omega) \\
   & &+  \sum_{\gamma = \uparrow, \downarrow} \int \frac{d^2 q}{(2 \pi)^2} V_d(\vec{q}) \hat{G}_{\alpha, \alpha}^{(o)}(\vec{K}, \vec{K}, \omega)
  \left[ \exp ( i \frac{\vec{q}}{2} \cdot \vec{A}(\vec{K}) ) \right]_{\alpha, \gamma} \hat{G}_{\gamma, \beta}(\vec{K} - \vec{q}, \vec{K}'; \omega)
\end{eqnarray}
 The matrix $\vec{A}(K)$ in eq.14 is defined in eq. 7.  $\hat{G}^{(o)}_{\alpha, \alpha}(\vec{K}, \vec{K}'; \omega)$ is the  Green's function in the absence of the  disorder potential and the external electric field. Using eq. 14, we perform an average over the disorder potential $V_d (\vec{q})$.  We introduce the averaged single particle Green's function, $\overline{G_{\alpha,\beta}(\vec{K}, \vec{K}'; \omega)}$ and compute the single particle occupation.We substitute this Green's function in eqs. 12 and 13, and consider here only the $non$ $dissipative$ $velocity$ given in eq.10. For this case the Hall  current is determined by  the $monopole$ $strength$ given in eq. 8.

For the charge-Hall current, we find
\begin{eqnarray}
J^{(C)}_{2,Hall}=\frac{e^{2}}{h}E^{ext}_{1}\int\frac{d\omega}{4\pi i}\int d^{2}K\delta^{2}(K)&[&\cos\theta(K)\left(\overline{G_{\uparrow,\uparrow}(\vec{K},\vec{K};\omega)}-\overline{G_{\downarrow,\downarrow}(\vec{K},\vec{K};\omega)}\right)\nonumber\\
&&-\sin\theta(K)\left(\overline{G_{\uparrow,\downarrow}(\vec{K},\vec{K};\omega)}-\overline{G_{\downarrow,\uparrow}(\vec{K},\vec{K};\omega)}\right)]
\label{Hall charge current}
\end{eqnarray}
The magnetic spin current is given by
\begin{equation}
J^{(M)}_{2,Hall}=\frac{e g \mu_{B}}{2 h} E^{ext}_{1}\int\frac{d\omega}{4\pi i}\int d^{2}K\delta^{2}(K)[\overline{G_{\uparrow,\uparrow}(\vec{K},\vec{K};\omega)}+\overline{G_{\downarrow,\downarrow}(\vec{K},\vec{K};\omega)}]
\label{Spin Hall Current}
\end{equation}
The two sets of equations represent the exact formulation of the currents in terms of the averaged single particle  Green's functions.The presence of the two dimensional delta function $\delta^{2}(K)$ in eqs. 15,16 gives rise to the Hall current induced by the external electric field.The origin of the delta function are the  non-commuting 
cartesian coordinates given in eq.5. 

\textit{Exact results for the Hall current}.  Eqs. 15 and 16 depend only on the single particle Green's function residue at $\vec{K}=0$. In the $absence$ of $disorder$, the residue is one and the $spin$ $Hall$ $conductance$ for a $zero$ $magnetic$ $field$ is $\sigma_{2, Hall}^{(M)} = \frac{e g \mu_B}{2 h}$ . This exact result is a reflection of the commutator given  in eq. 8.  In the units used in ref. 11, our result corresponds to $\sigma_{2, Hall}^{(M)} = \frac{e}{4 \pi}$ which is larger by a factor of two in comparison with the value given in ref.11. In ref. 11 the monopole has not been identified.  Instead one uses the equation of motion for a spin operator in the presence of an effective field computed within the mean field approximation.  We believe that this is the reason of the discrepancy between the two results.

\textit{The averaged  single particle Green's function }.  The averaged Green's function $\overline{G_{\alpha, \beta}(\vec{K}, \vec{K}'; \omega)}$ is given in terms of the self energy $\Sigma_{\alpha, \alpha}(\vec{K}, \omega)$.  Using eq. 14, we find $\overline{G_{\alpha, \beta}(\vec{K}, \vec{K}'; \omega)} = \delta_{\alpha,\beta} \delta_{\vec{K}, \vec{K}'} \left[(G_{\alpha, \alpha}^{(o)}(\vec{K}, \omega))^{-1} - \Sigma_{\alpha, \alpha}(\vec{K}, \omega) \right]^{-1} $.  Using $\overline{V_{d}(\vec{q})}=0$ and  $\overline{V_{d}(\vec{q})V_{d}(\vec{q'})} = W^{2}_{0} \delta_{\vec{q}, - \vec{q'}}$, where $W_{0}$  represents the effective potential, we find for a zero magnetic field in the limit $\vec{K} \rightarrow 0$ the self energy $\Sigma_{\uparrow, \uparrow}(\vec{K}, \omega) = \Sigma_{\downarrow, \downarrow}(\vec{K}, \omega) \equiv \Sigma_\parallel(\vec{K},\omega)$.  Therefore we have $\overline{G_{\uparrow,\uparrow}} = \overline{G_{\downarrow,\downarrow}} \equiv \overline{G_\parallel}$, $\overline{G_{\|}(\vec{K},\omega)} = \frac{\vartheta(|\vec{K}|-K_{F})}{\hbar\omega-\epsilon_{so}-\epsilon(\vec{K})- \Sigma_\parallel^{(R)}(\vec{K}, \omega)+\frac{i\hbar}{2\tau_{\|}}}+\frac{\vartheta(K_{F}-|\vec{K}|)}{\hbar\omega-\epsilon_{so}-\epsilon(\vec{K})- \Sigma_\parallel^{(R)}(\vec{K}, \omega) -\frac{i\hbar}{2\tau_{\|}}}$.
The Green's function is a function of the Fermi-Dirac step function $\vartheta$ at $T=0$.  At finite temperatures, we replace the step function with the thermal Fermi-Dirac occupation function $f_{F.D.}(\epsilon_{so}+\epsilon(K)+\Sigma_\parallel^{(R)}({K})-E_{F})$, which depends on the Fermi energy $E_{F}=\frac{\hbar^{2}K_{F}^{2}}{2m^{\ast}}$).  The imaginary part of the self energy allows us to compute the inverse of the life time $\tau_\parallel$ which is controlled by the effective elastic scattering strength $W^{2}_{0}$.  $\Sigma_\parallel^{(R)} (K,\omega)$ is the real part of the self energy $\Sigma_\parallel (K, \omega)$ .The real part of the self energy determines the residue of the Green's function  for $k=0$ .(In the presence of strong electron-electron interaction the residue might vanishes causing the Hall current to to be zero.)Due to disorder the real part of the self energy is given by $\Sigma_\parallel^{(R)}(\vec{K}, \omega) = P \int \frac{d^2 q}{(2 \pi)^2} \frac{|V_d(\vec{q})|^2}{\hbar \omega - \epsilon(\vec{K}-\vec{q})} \approx \frac{m \pi W_0^2}{h^2} P \int_0^{\hbar \Omega_{max}} \frac{d \epsilon}{\hbar \omega - \epsilon} = \frac{m \pi W_0^2}{h^{2}} \log (\frac{\omega}{\Omega_{max}})$, where $\hbar \Omega_{max}$ is the cutoff energy and $P$ stands for the principle value.  We observe that in the limit $\omega \rightarrow 0$ is problematic since it corresponds to a situation where the single particle wavelength is larger than the elastic mean free path and therefore our perturbative result in $momentum$ $space$ is not valid.This difficulty is  avoided for finite system and weak disorder where we find that the real part of the self energy is finite.The infinite limit has been considered in the absence of the spin orbit interaction within the real space $Locator$ expansion \cite{Kroha} one finds that the averaged self energy is finite in the limit $\omega\rightarrow0$.(We know from the localization theory  that the bottom of the spectrum is localized  ,this implies that the momentum representation used  for the self energy is not valid  in this limit.For this reason we will limit ourself to finite systems or infinite systems and no disoder.)

\textit{The Hall current }.  The spin Hall current for $B=0$ and $B \neq 0$ at $T=0$ for a  2DEG is given by $\sigma_{2, Hall}^{(M)} = \frac{e g \mu_B}{2 h}$.This result follows from the fact that $\Sigma_{\|}^R(K \rightarrow 0, \omega)$ is finite and the residue of the Green's function is one.  The charge Hall conductance vanishes for a zero magnetic field.  For a non-zero magnetic field $b_3 \neq 0$ ,the charge Hall current is derived from eq. 15.  We find the current,
\begin{equation}
J^{(C)}_{2,Hall}=E^{ext}_{1}\frac{e^{2}}{h}\frac{1}{2}\left[f_{F.D.}(\epsilon_{so}+\Sigma_{\|}^{(R)}(0)+b_{3}-E_{F})-f_{F.D.}(\epsilon_{so}+\Sigma_{\|}^{(R)}(0)-b_{3}-E_{F})\right] 
\label{Charge Hall Current}
\end{equation} 
  Using equation $17$ we find that  the charge Hall conductivity is given by $\sigma^{(C)}_{2,Hall}=\frac{e^{2}}{h}\exp\left[-\frac{E_{F}-\Sigma_{\|}^{(R)}(0)-\epsilon_{so}}{k_{B}T}\right]\sinh(\frac{g\mu_{B}B}{2 k_{B}T})$.  This result shows  that  the conductivity  increases with the magnetic field and vanishes at $T \rightarrow 0$. 

\textit{The  Hall conductivity for a system with a width $D_z$}.  We consider a situation where the two dimensional system of size $L\times L$ has a finite  width $D_{z}$ in the $i=3$ direction such that $L >> D_z$.  As a result, the single particle energy is given by $E(K,n)$ where $\vec{K}$ is a two dimensional vector and $n$ represents the quantization in the $i=3$ direction.  The single particle energy $\epsilon(K)$ is replaced by $E(K,n)=\frac{\hbar^{2}}{2m^{\ast}}\left[K^{2}+n^{2}(\frac{\pi}{D_{z}})^{2}\right]$ , $n = 0, 1, 2,\ldots$.  The spin orbit interaction given in equation $2$
is two dimensional (in agreement with the experimental situation )where  the intrinsic electric field is in the $i=3$ direction.  Our case is not valid for cases where the intrinsic electric field  has three components and gives rise to a three dimensional spin orbit interaction.  For such a case, we have to compute the Hall conductivity along the line used for the the Quantum Hall in three dimensions \cite {mont}.  For the remaining case we will consider only the case where the intrinsic electric field is in the $i=3$ direction.  Under this condition, the  Hall current is controlled by an external voltage $U_{G}$ applied in the $i=3$ direction.This voltage shifts  the Fermi energy $E_{F}$ to $E_{F,G}$,  $E_{F}\rightarrow E_{F,G}\equiv E_{F}-e U_{G}$.

The charge-Hall conductivity is computed for two cases :a) $k_{B}T<\frac{\hbar^{2}\pi^{2}}{2m^{\ast}D_{z}^{2}}$ for this case    the  conductance is nonzero only when the the resonance condition   ,$E(K=0,n)+\epsilon_{so}=E_{F,G}$  is achieved, $\sigma^{(C)}_{2,Hall}=\frac{e^{2}}{h}(\frac{g\mu_{B}B}{k_{B}T})\left[\frac{1}{4}\sum_{n=0}^{\infty} cosh^{-2}(\frac{E(K=0,n)+\epsilon_{so}-E_{F,G}}{2k_{B}T})\right]$. 

  b) In the oposite situation (level spacing smaller than the thermal energy) we find that the charge-Hall conductance is given by,
  $\sigma^{(C)}_{2,Hall}=\frac{e^{2}}{h}\frac{g\mu_{B}B}{\sqrt{E_{F,G}-\epsilon_{so}}}\sqrt{\frac{m^{\ast}}{2\hbar^{2}}}\frac{D_{z}}{\pi}$.
For this case we see that the charge -Hall conductivity increases with the external magnetic field $B$.

 The spin Hall conductivity  for a system with a width $D_{z}$ is computed  with the help of eq.16 and find ,  
\begin{eqnarray}
\sigma^{(M)}_{2,Hall}&=&\frac{e g \mu_{B}}{2h} \int \frac{d\omega}{4\pi} \sum_{n=0}^{\infty}\int d^{2}K\delta^{2}(K)\nonumber \\ & & f_{F.D.}(\hbar\omega-E_{F,G})[Im\frac{-2}{\hbar\omega-E(K,n)-\epsilon_{so}- \Sigma_{\|}^{(R)}(0) - E_{F,G}+i\frac{\hbar}{2\tau_{\|}}}]
\label{Spin-Hall conductance}
\end{eqnarray} 
When  the level spacing $\frac{\hbar^{2}}{2m^{\ast}} \frac{\pi^{2}}{D_{z}^{2}}$ is larger than the thermal energy $k_{B}T$ and the broadening energy $\frac{\hbar}{2\tau_{\|}}$, we obtain a set of plateaus for the Spin Hall conductance given in eq.18.  The value of the conductance depends on the effective chemical potential $ E_{F,G}\equiv E_{F}- e U_{G}$.  We compute the $residue$ for eq.18 and find that the spin-Hall conductance is quantized.  The Spin Hall conductance is quantized in units of $\frac{e g \mu_{B}}{2 h}$ and is given by $\sigma^{(M)}_{2,Hall} = N \frac{e g \mu_{B}}{2 h}$, where the integer $N$ is determined by the number of occupied bands $N \leq \left[ \sqrt{\frac{2m^{\ast}(E_{F,G} - \epsilon_{so}-\Sigma^R(0))}{\hbar^{2}}}\frac{D_{z}}{\pi}\right]\leq N+1$.

We propose an experiment to  measure the spin-Hall conductivity given given by eq.18.      We observe  that if we replace in eq. 2 the external electric field $E_1^{ext}$ by an external magnetic field gradient $H^{ext}_2$ in the $i=2$ direction the last term in equation $2$ is replaced by $\frac{1}{2}g \mu_{B}\sigma_{3}H^{ext}_{2}(\frac{i \partial}{\partial K_{2}})$.As a result the charge and spin  Hall currents will be interchanged.Therefore a  charge Hall current will be driven by a $magnetic$ $field$ $gradient$. The charge Hall conductivity  will be given for this case  by the spin Hall conductivity computed in equation 18.  We find that  an electric current $I^{(C)}_{1}$ will be generated in the $i=1$ direction, if a magnetic field is applied in the $i=3$ direction with a gradient $H^{ext}_{2}$ in the $i=2$ direction.  Under this condition, a charge Hall current wil be given by the relation $I^{(C)}_{1} = N \left(\frac{eg\mu_{B}}{2 h}\right)H^{ext}_{2}\approx N(1.5\cdot 10^{-9})H^{ext}_{2}$.
For N=1 and a magnetic field gradient $H^{ext}_{2}$ of $1$ $Tesla/meter$, our formula predicts  a current of $1.5 \cdot 10^{-9}$ Amperes, which should be observed experimentally at temperatures $k_{B} T \leq \frac{\hbar}{2\tau_{\|}}$.  

To conclude, we have shown that the $SU(2)$ transformation used to diagonalize the spin orbit problem induces no-commutative coordinates and gives rise to a \textit{non-dissipative Hall-current}.  The spin-Hall conductance is quantized in units of $\frac{e g \mu_{B}}{2 h}$  which might be observed by using a magnetic field gradient.  

 \center { FIGURE}

FIG.$1$ shows  the Hall current induced by the magnetic field in the $i=3$ direction with a gradient in the $i=2$ direction ,$H^{ext}_{2}$.The induced charge current $I_{c}$is in the $i=1$ direction and is proportional to the spin Hall conductance.

\end{document}